\documentclass[11pt]{article}
\usepackage{amsfonts}
\usepackage{axodraw}
\def\eref#1{(\ref{#1})}
\def\pol{\epsilon}
\def\zb{\bar{z}}
\def\sumYM{\sum_{S_1, S_2 \in \{N=4\} }}
\def\sumGrav{\sum_{S_1, S_2 \in \{N=8\} }}
\def\lr{\leftrightarrow}

\def\twoloop{{2 \mbox{-} \rm loop}}
\def\e{\epsilon}
\def\Tr{\, {\rm Tr}}
\begin{document}
\begin{titlepage}
\begin{flushright}

hep-th/9911194 \hfill SLAC--PUB--8303\\ 
UCLA/99/TEP/36\\
SWAT-99-243\\
November, 1999\\
\end{flushright}

\vskip 1.3cm

\begin{center}
{\Large\bf Perturbative Relations between Gravity and Gauge Theory$^\star$}\\ 

\vspace{1.0cm}

{Zvi Bern$^1$}\\
\vspace{.2cm}
{\sl Department of Physics\\
University of California at Los Angeles\\
Los Angeles, CA 90095-1547, USA}\\

\vspace{.5cm}

{Lance Dixon$^2$ and Maxim Perelstein$^2$}\\
\vspace{.2cm}
{\sl Stanford Linear Accelerator Center\\ 
Stanford University\\ 
Stanford, CA 94309, USA}\\

\vspace{.5cm}

{David C. Dunbar$^3$}\\
\vspace{.2cm}
{\sl Department of Physics\\
University of Wales Swansea\\ 
Swansea, SA2 8PP, UK}\\

\vspace{.5cm}

{Joel S. Rozowsky$^4$}\\
\vspace{.2cm}
{\sl Institute for Fundamental Theory\\ 
Department of Physics\\
University of Florida\\
Gainesville, FL 32611}\\
\vspace{.5cm}
\end{center}

\begin{abstract}
We review the relations that have been found between multi-loop scattering
amplitudes in gauge theory and gravity, and their implications for
ultraviolet divergences in supergravity.
\end{abstract}

%PACS numbers: 04.50, 04.65,+e, 11.15.Bt

\vspace{0.9cm}
\begin{center}
{\sl Submitted to Classical and Quantum Gravity}
\end{center}

\vfill
\noindent\hrule width 3.6in\hfil\break
${}^1$Research supported in part by the US Department of Energy 
under grant DE-FG03-91ER40662.\hfil\break
${}^2$Research supported by the US Department of Energy 
under grant DE-AC03-76SF00515.\hfil\break
${}^3$Research supported in part by the Leverhulme Foundation.\hfil\break
${}^4$Research supported by the US Department of Energy 
under grant DE-FG02-97ER41029.\hfil\break
${}^\star$ Presented by L.D. at {\sl Strings 99}, July 19-25, 1999, 
Potsdam, Germany.\hfil\break
\end{titlepage}

\baselineskip 16pt

%%%%%%%%%%%%%%%%%%%%%%%%%%%%%%%%%%%%%%%
\section{Introduction}

Although Yang-Mills theory and Einstein gravity are both based on
local symmetries, and include long-range forces at the classical level,
they are very different theories.  For example, Yang-Mills theory is
renormalizable, and due to asymptotic freedom, highly nontrivial in the
infrared region; whereas gravity is nonrenormalizable, and its ultraviolet
structure is of the greatest interest theoretically.  There are also
great differences between the perturbative Feynman rules for the two
theories:  Those for Yang-Mills theory contain only three- and four-point 
vertices, while those for gravity can have arbitrarily many external legs.
Nevertheless, string theory suggests the heuristic relationship,
\begin{equation}
\hbox{gravity} \sim \hbox{(gauge theory)} \times \hbox{(gauge theory)} \,,
\label{GravityYMRelation}
\end{equation}
following from the representation of string amplitudes as integrals over
world-sheet variables --- complex integrals for closed strings (gravity)
and real integrals for open strings (gauge theory) --- and the
factorization of the closed-string integrand into two copies of the
open-string integrand.  This relationship was made precise by 
Kawai, Lewellen and Tye (KLT)~\cite{KLT} for tree-level scattering
amplitudes.  In this talk we describe the relations between multi-loop 
scattering amplitudes in gravity and gauge theory that were obtained in 
ref.~\cite{BDDPR}, and discuss some of their implications.

Multi-loop scattering amplitudes are of considerable interest in 
both nonabelian gauge theory and gravity (and their supersymmetric 
extensions).  On the gauge theory side, there are practical implications
for more precise predictions of jet rates and other QCD phenomena observed 
in collider experiments.  To date, no jet rate computations have been
carried out beyond next-to-leading order in the strong coupling
$\alpha_s$, even though in many cases, such as $e^+e^-$ annihilation into
three jets, experiment demands one higher order.   Two-loop scattering
amplitudes are required to calculate such next-to-next-to-leading order
corrections.

On the gravity side, ultraviolet properties are of primary interest.
Although gravity is nonrenormalizable by power counting, and the
conventional wisdom is that only string theory could possibly render it
finite to all orders, in point of fact no divergence has yet been
established for any supersymmetric theory of gravity in four dimensions.
Nonsupersymmetric theories of gravity with matter generically diverge at
one loop~\cite{tHooftVeltmanAnnPoin,tHooftGrav,DeserEtal}, and pure
gravity diverges at two
loops~\cite{PureGravityInfinityGSV}.  However,
supersymmetry Ward identities (SWI)~\cite{SWI} forbid all possible
one-loop~\cite{OneLoopSUGRA} and two-loop~\cite{Grisaru} counterterms in
any supergravity theory in $D=4$.  Thus at least a three-loop calculation is
required to definitively answer the question of the finiteness of
four-dimensional supergravity.  In addition, there is a candidate
counterterm at three loops for all supergravities including the maximally
extended version ($N=8$)~\cite{KalloshNeight}.  In ref.~\cite{BDDPR},
divergences were computed in (higher-dimensional) $N=8$ supergravity, but
only up to two loops.  We expect, though, that the same techniques should
be applicable beyond the two-loop level.  In fact, our work suggests a 
natural conjecture for the divergences appearing at $L$ loops (see
section~\ref{GravDiv}).  The conjecture would imply that $N=8$ 
supergravity in $D=4$ first diverges at five loops, not three loops.

Loop amplitudes in $D=11$ supergravity can also be used
to extract information about M theory~\cite{StringR4,GreenTwoLoop}.
Here the finite parts of the amplitudes are most important, particularly 
their dependence on the radii of one or two compactified dimensions.

In principle one could investigate higher-loop relations between gravity
and gauge theory in superstring theory using the world-sheet
representation.  However, at the multi-loop level this becomes technically
rather difficult.  Also, we would like to study the ultraviolet properties
of particle theories of gravity.  Recently, the KLT relations have been
examined at the Lagrangian level, by introducing an auxiliary scalar field
(i.e. the dilaton) into the Einstein-Hilbert action and carrying out
appropriate field redefinitions~\cite{BernGrant}.  However, we won't
pursue this direction further here.  Instead we will use unitarity as a tool.
Unitarity has proven very useful for one-loop QCD
computations~\cite{Review}, and it has also been applied to two-loop $N=4$
super-Yang-Mills amplitudes~\cite{BRY}.  The basic approach is to
calculate the unitarity cuts of an amplitude, and then find functions that
reproduce all such cuts.  One can construct a perturbative bootstrap from
tree amplitudes up to multi-loop amplitudes in this way.  In the case of
gravity, where the KLT relations express its tree amplitudes as roughly
`squares' of those of gauge theory, unitarity is particularly effective:
The same algebraic steps employed in simplifying the gauge theory cuts can
often be recycled in the (otherwise more complex) corresponding gravity
cuts.

%%%%%%%%%%%%%%%%%%%%%%%%%%%%%%%%%%%%%%%%%%%%%%
\section{Tree-level relations}

The starting point for investigating gravity--gauge theory relations
via unitarity is the set of tree-level relations found by KLT~\cite{KLT}.
The key observation is that any closed string vertex
operator is a product of open string vertex operators,
\begin{equation}
V^{\rm closed} = V_{\rm left}^{\rm open}\, \times \,  
\overline{V}_{\rm right}^{\rm  open} \,.
\label{ClosedVertex}
\end{equation}
This product structure is then reflected in the amplitudes, written
as correlation functions of vertex operators.

For example, the open string tree amplitude for $n$ gluons is
\begin{equation} 
A_n \sim \int {dx_1 \cdots d x_n \over   {\cal V}_{abc} }
\prod_{1  \le i < j \le n}  |x_ i - x_j|^{k_i \cdot k_j}
\exp\biggl[ \sum_{i <j} \Bigl({\pol_i \cdot \pol_j \over (x_i - x_j)^2}
                       + {k_i \cdot \pol_j - k_j \cdot \pol_i
                          \over (x_i - x_j)} \Bigr) \biggr] \,
                        \biggr|_{\rm lin.} \,,
\label{OpenKN}
\end{equation}
where
\begin{equation}
{\cal V}_{abc} = {dx_a\, dx_b\, dx_c \over |(x_a - x_b)
     (x_b - x_c) (x_c-x_a)|} \, , 
\end{equation}
and $x_a,x_b,x_c$ are any three of the $x_i$.  In equation~\eref{OpenKN}
we have suppressed the inverse string tension $\alpha'$, and 
the `lin.' denotes that after expanding the exponential one only
keeps terms linear in each polarization vector $\pol_i$.
The corresponding closed-string $n$-graviton amplitude is
\begin{eqnarray}
M_n &\sim& \int {d^2z_1 \cdots d^2 z_n \over  \Delta_{abc} }
\prod_{1  \le i < j \le n}  (z_ i - z_j)^{ k_i \cdot k_j}
\exp\biggl[ \sum_{i <j} \Bigl({\pol_i \cdot \pol_j \over (z_i - z_j)^2}
                       + {k_i \cdot \pol_j - k_j \cdot \pol_i
                          \over (z_i - z_j)} \Bigr) \biggr] 
\nonumber\\
&& \hskip2.3cm \times
\prod_{1  \le i < j \le n}  (\zb_ i - \zb_j)^{ k_i \cdot k_j}
\exp\biggl[ \sum_{i <j} \Bigl({\overline{\pol}_i \cdot 
\overline{\pol}_j \over (\zb_i - \zb_j)^2}
+ {k_i \cdot \overline{\pol}_j - k_j \cdot \overline{\pol}_i
                          \over (\zb_i - \zb_j)} \Bigr) \biggr] 
          \Bigr|_{\rm lin.} \,, 
\label{ClosedKN}
\end{eqnarray}
where
\begin{equation}
\Delta_{abc} = {d^2z_a \, d^2 z_b \, d^2 z_c \over
            |z_a - z_b|^2 |z_b - z_c|^2 |z_c-z_a|^2 } \,,
\end{equation}
and $z_a,z_b,z_c$ are any three of the $z_i$.
In a helicity basis~\cite{SpinorHelicity}, one can write
the graviton polarization tensor as a product of gluon
polarization vectors, 
$\pol^{\mu\nu}_i(\pm) 
= \pol^\mu_i(\pm) \, \overline{\pol}^\nu_i(\pm)$.

The closed string integrand in~\eref{ClosedKN} is a product of two
open string integrands. This factorization holds for general closed-string
states, not just gravitons, using the tensor product decomposition of
closed string states in terms of open string ones, so that the vertex
operator relation~\eref{ClosedVertex} can be applied.

From equations~\eref{OpenKN} and~\eref{ClosedKN}, various 
contour-integral deformations lead
to relations between tree-level closed and open string amplitudes
after all integrations have been performed~\cite{KLT}.  
Here we will need only the KLT relations in the limit $\alpha' \to 0$ 
for the four- and five-point amplitudes~\cite{BGK}:
\begin{eqnarray}
M_4^{\rm tree} (1,2,3,4) &=&  
  - i s_{12} A_4^{\rm tree} (1,2,3,4) \, A_4^{\rm tree}(1,2,4,3)\,, 
\label{GravYMFour}\\
M_5^{\rm tree}(1,2,3,4,5) &=& i s_{12} s_{34}  A_5^{\rm tree}(1,2,3,4,5)
                                     A_5^{\rm tree}(2,1,4,3,5) \nonumber\\
&& \hskip0.5cm 
 + i s_{13}s_{24} A_5^{\rm tree}(1,3,2,4,5) \, A_5^{\rm tree}(3,1,4,2,5) \,,
\label{GravYMFive}
\end{eqnarray}
where the $M_n$'s are the amplitudes in a gravity theory stripped of 
couplings, the $A_n$'s are the color-ordered subamplitudes in a gauge
theory and $s_{ij}\equiv (k_i+k_j)^2$.  We suppress all
$\pol_j$ polarizations and $k_j$ momenta, but keep the `$j$'
labels to distinguish the external legs.  
The full amplitudes are given by
\begin{eqnarray}
{\cal M}_n^{\rm tree}(1,2,\ldots n) &=& 
\left({  \kappa \over 2} \right)^{(n-2)} 
M_n^{\rm tree}(1,2,\ldots n)\,,
\nonumber\\
{\cal A}_n^{\rm tree}(1,2,\ldots n) &=& g^{(n-2)} \sum_{\sigma \in S_n/Z_n}
{\rm Tr}\left( T^{a_{\sigma(1)}} 
T^{a_{\sigma(2)} }\cdots  T^{a_{\sigma(n)}} \right)
 A_n^{\rm tree}(\sigma(1), \sigma(2),\ldots, \sigma(n)) \,,
\label{fullamps}
\end{eqnarray}
where $S_n/Z_n$ is the set of all permutations, but with cyclic
rotations removed, $g$ is the gauge theory coupling constant, and
$\kappa^2 = 32\pi G_N$.  The $T^{a_i}$ are fundamental representation 
matrices for the Yang-Mills gauge group $SU(N_c)$, normalized so that
$\Tr(T^aT^b) = \delta^{ab}$.  

Relations~\eref{GravYMFour} and \eref{GravYMFive} can also be used when
each external graviton state in $M_n$ is replaced by any of the 256
states of the $N=8$ supergravity multiplet.  The Fock space decomposition
\begin{equation}
\vert N=8\ \hbox{SUGRA state} \rangle
\ =\ \vert N=4\ \hbox{super YM state} \rangle 
\otimes \vert N=4\ \hbox{super YM state} \rangle 
\end{equation}
should then be used to select the corresponding states for the two 
$A_n$ factors on the right-hand side of the relation.

%%%%%%%%%%%%%%%%%%%%%%%%%%%%%%%%%%%%%%%%%%%%%%%%%%%
\section{Loop amplitudes from unitarity}

Unitarity of the $S$ matrix, $S^\dag S = 1$, written in terms of the $T$ 
matrix defined by $S \equiv 1 + i T$, reads
\begin{equation}
 2 \, {\rm Im} T_{if}\ =\ \sum_j T^*_{ij} T_{jf} \, , 
\label{BasicUnitarity}
\end{equation}
where $i$ and $f$ are initial and final states, and the `sum' is over
intermediate states $j$ (and includes an integral over intermediate
on-shell momenta).  Perturbative unitarity consists of expanding both
sides of equation~\eref{BasicUnitarity} in terms of coupling constants,
$g$ for gauge theory and $\kappa$ for gravity, and collecting terms of the
same order.  For example, the imaginary (or absorptive) parts of one-loop
four-point gauge amplitudes (order $g^4$) are given in terms of the 
product of two four-point tree amplitudes ($g^2 \times g^2$).
The cuts in two-loop four-point gauge amplitudes (order $g^6$)
are of two types:  the product of a four-point tree amplitude and a 
four-point one-loop amplitude ($g^2 \times g^4$), and the product of
two five-point tree amplitudes ($g^3 \times g^3$).  In terms of the 
number of particles crossing the cut, the former is a two-particle cut,
the latter a three-particle cut.

%%%%%%%%%%%%%%%%
\subsection{Two-particle cutting equations}

The two-particle cuts in $N=4$ super-Yang-Mills theory have a very simple
self-replicating structure~\cite{BRY}.  The key equation is
\begin{equation}
\sumYM A_4^{\rm tree}(-\ell_1^{S_1}, 1, 2, \ell_2^{S_2}) \times
  A_4^{\rm tree}(-\ell_2^{S_2}, 3, 4, \ell_1^{S_1})
= - i { s_{12} s_{23} \over (\ell_1 - k_1)^2 \, (\ell_2 - k_3)^2 } 
    A_4^{\rm tree}(1,2,3,4) \,, 
\label{BasicYMCutting}
\end{equation}
where $\ell_{1,2}$ are the intermediate momenta, and $S_{1,2}$ label 
states of the $N=4$ multiplet.  The $N=4$ labels corresponding to the 
external states with momenta $k_i$ have been suppressed, but 
equation~\eref{BasicYMCutting} is valid for arbitrary combinations of
external states.  It is also valid for arbitrary (not just 
four-dimensional) momenta.  One way to derive 
the equation is by working `backwards' from the one-loop
$N=4$ amplitudes first obtained using string theory~\cite{GSB}.

Equation~\eref{BasicYMCutting} can be represented graphically as
\begin{equation}
\begin{picture}(300,60)(0,0)
\SetWidth{1.0}
\Text(2,30)[r]{{\Huge $\Sigma$}} \Text(3,10)[r]{{\tiny $N=4$}}
\Text(18,10)[r]{$1$}  \Text(18,50)[r]{$2$}
\Line(20,10)(30,20)   \Line(20,50)(30,40)
\Line(40,20)(50,10)   \Line(40,40)(50,50)
\Line(53,10)(63,20)   \Line(53,50)(63,40)
\Line(73,20)(83,10)   \Line(73,40)(83,50)
\Text(85,50)[l]{$3$}  \Text(85,10)[l]{$4$} 
\GOval(35,30)(20,7)(0){0.5}
\GOval(68,30)(20,7)(0){0.5}
\Text(100,30)[l]{$=\ -i s_{12} \, s_{23}$}
\Text(168,10)[r]{$1$}  \Text(168,50)[r]{$2$}
\Line(170,10)(180,20)   \Line(170,50)(180,40)
\Line(190,20)(200,10)   \Line(190,40)(200,50)
\Text(202,50)[l]{$3$}  \Text(202,10)[l]{$4$} 
\GOval(185,30)(20,7)(0){0.5}
\Line(220,10)(230,20)   \Line(220,50)(230,40)  \Line(230,20)(230,40)
\Line(230,20)(238,20)   \Line(230,40)(238,40)
\Line(242,20)(250,20)   \Line(242,40)(250,40)  \Line(250,20)(250,40)
\Line(250,20)(260,10)   \Line(250,40)(260,50)
\Text(265,20)[l]{$,$} 
\SetColor{Blue} 
\DashLine(51.5,0)(51.5,60){3}  \DashLine(240,0)(240,60){3}
\end{picture}
\label{BasicYMCuttingGraphical}
\end{equation}
where the blobs represent tree amplitudes, and the two denominator factors
in~\eref{BasicYMCutting} are represented kinematically by the 
two internal propagators in the $\phi^3$ diagram on the far right.
This representation makes clear that the one-loop two-particle cut is
given by a cut scalar box integral, multiplied by the tree amplitude
and a simple overall factor.  The full one-loop amplitude is obtained
simply by replacing the cut scalar box integral with the full scalar 
box integral ($D$ is the spacetime dimension),
\begin{equation}
\begin{picture}(40,40)(0,17)
\SetWidth{1.0}
\Line(0,0)(10,10)   \Line(0,40)(10,30)  \Line(10,10)(10,30)
\Line(10,10)(30,10)   \Line(10,30)(30,30)
\Line(30,10)(30,30) \Line(30,10)(40,0)   \Line(30,30)(40,40)
\end{picture}
\ \equiv \int {d^D\ell\over(2\pi)^D} { 1 \over \ell^2 (\ell-k_1)^2
                                   (\ell-k_1-k_2)^2 (\ell+k_4)^2 } \,.
\label{boxintdef}
\end{equation}
\vskip 0.5 cm

%%%%%%%%%%%%%%%%%%%
\subsection{$N=4$ iteration \label{IterationSection}}

Because the dependence of the two-particle cut on the external $N=4$ 
states is just that of the tree amplitude, the two-particle cuts can
be iterated easily.  For example, the two-loop two-particle cut is 
given by
\begin{eqnarray}
\sum_{N=4}
\begin{picture}(100,60)(0,30)
\SetWidth{1.0}
\Text(18,10)[r]{$1$}  \Text(18,50)[r]{$2$}
\Line(20,10)(30,20)   \Line(20,50)(30,40)
\Line(40,20)(50,10)   \Line(40,40)(50,50)
\Line(53,10)(63,20)   \Line(53,50)(63,40)
\Line(83,20)(93,10)   \Line(83,40)(93,50)
\Text(95,50)[l]{$3$}  \Text(95,10)[l]{$4$} 
\GOval(35,30)(20,7)(0){0.5}
\GCirc(73,30){14.142}{0.5}
\GCirc(73,30){7.071}{1}
\SetColor{Blue}  \DashLine(51.5,0)(51.5,60){3} 
\end{picture}
&=& \sum_{N=4} 
\begin{picture}(100,60)(0,30)
\SetWidth{1.0}
\Text(18,10)[r]{$1$}  \Text(18,50)[r]{$2$}
\Line(20,10)(30,20)   \Line(20,50)(30,40)
\Line(40,20)(50,10)   \Line(40,40)(50,50)
\Line(53,10)(63,20)   \Line(53,50)(63,40)
\Line(73,20)(83,10)   \Line(73,40)(83,50)
\Text(85,50)[l]{$3$}  \Text(85,10)[l]{$4$} 
\GOval(35,30)(20,7)(0){0.5}
\GOval(68,30)(20,7)(0){0.5}
\SetColor{Blue}  \DashLine(51.5,0)(51.5,60){3} 
\end{picture}
\ \times (-i) s_{34} (\ell_2-k_3)^2\ \ 
\begin{picture}(40,60)(0,27)
\SetWidth{1.0}
\Line(0,20)(40,20)  \Line(0,40)(40,40)
\Line(10,20)(10,40) \Line(30,20)(30,40) 
\end{picture}
\nonumber\\
&=& 
\begin{picture}(60,60)(0,27)
\SetWidth{1.0}
\Text(18,10)[r]{$1$}  \Text(18,50)[r]{$2$}
\Line(20,10)(30,20)   \Line(20,50)(30,40)
\Line(40,20)(50,10)   \Line(40,40)(50,50)
\Text(52,50)[l]{$3$}  \Text(52,10)[l]{$4$} 
\GOval(35,30)(20,7)(0){0.5}
\end{picture}
\times  { (-i) s_{12} s_{23} \over (\ell_1-k_1)^2 (\ell_2-k_3)^2 } 
  \ (-i) s_{34} (\ell_2-k_3)^2\ 
\begin{picture}(60,60)(0,30)
\SetWidth{1.0}
\Line(0,20)(40,20)  \Line(0,40)(40,40)
\Line(10,20)(10,40) \Line(30,20)(30,40) 
\end{picture}
\nonumber\\
&=& - s_{12}^2 s_{23} 
    \times 
\begin{picture}(60,60)(0,27)
\SetWidth{1.0}
\Text(18,10)[r]{$1$}  \Text(18,50)[r]{$2$}
\Line(20,10)(30,20)   \Line(20,50)(30,40)
\Line(40,20)(50,10)   \Line(40,40)(50,50)
\Text(52,50)[l]{$3$}  \Text(52,10)[l]{$4$} 
\GOval(35,30)(20,7)(0){0.5}
\end{picture}
    \times
\begin{picture}(60,60)(0,27)
\SetWidth{1.0}
\Line(0,20)(18,20)  \Line(0,40)(18,40) \Line(10,20)(10,40)
\Line(22,20)(60,20)  \Line(22,40)(60,40)
\Line(30,20)(30,40) \Line(50,20)(50,40) 
\SetColor{Blue}  \DashLine(20,0)(20,60){3} 
\end{picture}
\ . \label{twoloopYMcut} 
\end{eqnarray}
\vskip 0.7 cm
Clearly, the coefficients of all multi-loop ladder diagrams can 
be determined in this way; but so can the `entirely two-particle 
constructible' diagrams, namely those which can be reduced to trees by
successive two-particle cuts, for example
\begin{equation}
\begin{picture}(60,40)(0,10)
\SetWidth{1.0}
\Line(0,0)(60,0)   \Line(0,40)(60,40)  
\Line(10,0)(10,40)  \Line(20,0)(20,40)  \Line(30,0)(30,40)  
\Line(50,0)(50,40)  \Line(30,10)(50,10) \Line(30,30)(50,30)
\Line(40,10)(40,30) \Line(40,20)(50,20)
\end{picture}
\ . \label{MondrianDiagram}
\end{equation}
\vskip 0.7 cm
In general, each two-particle cut through a channel with momentum 
$\ell_i+\ell_j$ results in an additional factor of $(\ell_i+\ell_j)^2$ 
multiplying the $\phi^3$ integral~\cite{BRY}.

At two loops, all terms in the $N=4$ super-Yang-Mills four-point amplitude
are detectable by the iterated two-particle cuts.  In order to confirm
that these are the only terms, one must calculate the more complicated 
three-particle cuts~\cite{BRY}.  Beginning at three loops (for nonplanar 
contributions), there are terms in the amplitude which do not have any
two-particle cuts at all, so three-particle cuts are required just to
guess their form.
  
The above discussion has been for color-ordered subamplitudes, which
need to be dressed with appropriate color factors to produce the full
gauge theory amplitude.  For the entirely two-particle constructible
terms, however, this dressing is very simple to describe:  In a graphical 
representation of the color factors, where a Kronecker $\delta^{ab}$ is
represented by an internal line and a structure constant $f^{abc}$ by
a three-vertex, one should multiply each kinematic ($\phi^3$) graph by
exactly the same color-factor graph.

%%%%%%%%%%%%%%%%%%
\subsection{Recycling gauge theory into gravity}

The first step in repeating the above analysis for $N=8$ supergravity
is to derive the corresponding two-particle cutting equation.
Using the four-point KLT relation~\eref{GravYMFour}, followed by the 
Yang-Mills cutting equation~\eref{BasicYMCutting},
the appropriate product of two gravity amplitudes is
\begin{eqnarray}
&&\sumGrav
M_4^{\rm tree}(-\ell_1^{S_1},  1, 2, \ell_2^{S_2}) \times
M_4^{\rm tree}(-\ell_2^{S_2}, 3, 4, \ell_1^{S_1}) \nonumber\\
&&\hskip2cm =  
- s_{12}^2 
\biggl( \sumYM A_4^{\rm tree}(-\ell_1^{S_1},  1, 2, \ell_2^{S_2}) \times
               A_4^{\rm tree}(-\ell_2^{S_2}, 3, 4, \ell_1^{S_1}) \biggr)
  \nonumber\\
&&\hskip2cm \qquad \quad \times     
\biggl( \sumYM A_4^{\rm tree}(\ell_2^{S_2},  1, 2, -\ell_1^{S_1}) \times
               A_4^{\rm tree}(\ell_1^{S_1}, 3, 4, -\ell_2^{S_2}) \biggr) 
\label{BasicGravityCutting} \\
&&\hskip2cm = \Bigl( s_{12} s_{23} A_4^{\rm tree}(1, 2, 3, 4) \Bigr)^2
   { s_{12}^2 \over (\ell_1 - k_1)^2  (\ell_2 - k_3)^2 
 (\ell_2 + k_1)^2  (\ell_1 + k_3)^2 } \nonumber\\
&&\hskip2cm= \Bigl( s_{12} s_{23} A_4^{\rm tree}(1, 2, 3, 4) \Bigr)^2
\biggl[{1\over (\ell_1 - k_1)^2 } + {1\over (\ell_1 - k_2)^2} \biggr]
\biggl[{1\over (\ell_2 - k_3)^2 } + {1\over (\ell_2 - k_4)^2} \biggr]
\, . \nonumber
\end{eqnarray}
In the last step we have performed a partial-fractioning of the
denominators (using on-shell relations), in order to get a form which is
recognizable as a sum of four different cut scalar box integrals,
corresponding to $1\lr2$ and $3\lr4$ permutations of the integral 
in~\eref{boxintdef}.

Equation~\eref{BasicGravityCutting} can be iterated whenever
equation~\eref{BasicYMCutting} can.  Its structure implies that the
coefficients of the corresponding $\phi^3$ integrals in $N=8$ supergravity
are essentially the squares of those in $N=4$ super-Yang-Mills theory
(once color factors have been removed from the latter)~\cite{BDDPR}.
Again the two-particle cuts should be checked via three- (and higher-)
particle cuts.  The check for gravity at two loops is greatly 
simplified by using the five-point KLT relation~\eref{GravYMFive}.  
The resulting two-loop $N=8$ supergravity amplitude is given by
\begin{eqnarray}
{\cal M}_4^{\twoloop}(1,2,3,4) &=& 
 -i \Bigl({\kappa \over 2} \Bigr)^6 
 [s_{12} s_{23} \,  A_4^{\rm tree}(1,2,3,4)]^2 \label{TwoLoopGrSquare}\\
&&\hskip0.5cm \times
\Biggl[ s_{12}^2 \biggl( \ \ 
\begin{picture}(70,40)(0,27)
\SetWidth{1.0}
\Text(3,20)[r]{1} \Text(3,40)[r]{2} 
\Line(5,20)(65,20)  \Line(5,40)(65,40) \Line(15,20)(15,40)
\Line(35,20)(35,40) \Line(55,20)(55,40) 
\Text(67,40)[l]{3} \Text(67,20)[l]{4} 
\end{picture}
\ \ + \ \ 
\begin{picture}(65,40)(0,27)
\SetWidth{1.0}
\Text(3,20)[r]{1} \Text(3,40)[r]{2} 
\Line(5,20)(45,20)  \Line(5,40)(45,40) \Line(15,20)(15,40)
\Line(35,30)(45,20) \Line(35,30)(45,40) 
\Line(55,30)(45,20) \Line(55,30)(45,40) 
\Line(30,30)(35,30) \Line(55,30)(60,30) 
 \Text(28,30)[r]{3}  \Text(62,30)[l]{4} 
\end{picture}
\ \ \biggr) + {\cal P}(2,3,4) \Biggr] \,, \nonumber
\end{eqnarray}
where `$+ {\cal P}(2,3,4)$' instructs one to add the five nontrivial
permutations of legs $2,3,4$.

Beyond two loops, complete results are not yet available; a full analysis
of higher-particle cuts must still be performed.

%%%%%%%%%%%%%%%%%%%%%%%%%%%%%%%%%%%%%%%%%%%%%%%%%%%%%%
\section{Ultraviolet divergences in $N=8$ supergravity \label{GravDiv}}

%%%%%%%%%%%%%%%%%%%%%%%%%%%%
\subsection{Two loops}

Two-loop ultraviolet divergences in $N=8$ supergravity can be extracted 
directly from the two-loop scattering amplitude~\eref{TwoLoopGrSquare},
by evaluating the divergences of the two $\phi^3$ double-box integrals 
that appear.  Since each integral has 7 propagators and $2D$ loop momenta
in the integration measure, ${\cal M}_4^{\twoloop}(1,2,3,4)$ is manifestly
finite for $D<7$.  This behavior is better than what was predicted by
power counting in an $N=4$ superspace formalism, which suggested that 
$N=8$ supergravity should diverge at two loops in $D=5$ and 
6~\cite{HoweStelleTownsend}.  The discrepancy is presumably due to the 
lack of a manifest $N=8$ invariance in the power-counting argument.
In the direct cut-based calculation, the on-shell $N=8$ supersymmetric 
Ward identities are utilized in summing over the intermediate states
crossing the cuts.

For $D\geq7$, no cancellations occur, and there are divergences in every
dimension of interest.  The results are~\cite{BDDPR},
\begin{equation}
  {\cal M}_4^{\twoloop,\ D=n-2\e}\vert_{\rm pole} 
  = {C_n\over\e\ (4\pi)^n} \times \left( {\kappa \over 2}\right)^6 
                         \times stu M_4^{\rm tree} \, ,
\label{gravtwolooppolesA}
\end{equation}
where
\begin{eqnarray}
C_7 &=&  {1\over2} \, {\pi\over3} (s^2+t^2+u^2) \,, \nonumber\\
C_9 &=&  {1\over4} \, {-13\pi\over 9\,072} (s^2+t^2+u^2)^2 \,, 
\label{gravtwolooppolesB}\\
C_{10} &=& {1\over12} \, { - 13 \over 25\,920 } \, stu \, 
  \bigl( s^2+t^2+u^2 \bigr) \,, \nonumber\\ 
C_{11} &=& {1\over48} \, {\pi\over5\,791\,500} 
  \Bigl( 438 (s^6+t^6+u^6) - 53 s^2 t^2 u^2 \Bigr) \,, \nonumber
\end{eqnarray}
and $s=s_{12}$, $t=s_{23}$, $u=s_{13}$.  We omit the $D=8$ two-loop
divergence because the $D=8$ theory diverges already at one loop, as can
be seen easily by inspecting the box integral~\eref{boxintdef}.  New
counterterms appear at two loops, though, as one would expect from a
nonrenormalizable theory.  (There are no corresponding one-loop
divergences in $D=9$ or $D=11$, in dimensional regularization, because all
invariants have even dimension.  In $D=10$, there is a cancellation of
one-loop divergences between the three different box integrals, because
$s+t+u=0$.)

The presence of a factor of $stu \, M_4^{\rm tree}$ in 
equation~\eref{gravtwolooppolesA} implies that in all the above cases,
for four graviton external states, the linearized counterterms take the 
form of derivatives acting on the operator
\begin{equation}
t_8 t_8 R^4 \equiv
t_8^{\mu_1\mu_2\cdots \mu_8}\, 
t_8^{\nu_1\nu_2\cdots \nu_8} \, 
R_{\mu_1\mu_2 \nu_1 \nu_2} \, 
R_{\mu_3\mu_4 \nu_3 \nu_4} \,
R_{\mu_5\mu_6 \nu_5 \nu_6} \,
R_{\mu_7\mu_8 \nu_7 \nu_8}  \,,
\label{Rfourterms}
\end{equation}
where the tensor $t_8$ is defined in equation (9.A.18) of ref.~\cite{GSW},
plus the appropriate $N=8$ completion.  When the indices are restricted to
four dimensions, $t_8t_8R^4$ becomes equal to the Bel-Robinson 
tensor~\cite{BelRobinson}.  We note that the 
operator~\eref{Rfourterms} appears in the tree-level superstring
effective action~\cite{GrossWitten}.  It also appears as the
one-loop counterterm for $N=8$ supergravity in $D=8$.  Finally, by
calculating the same amplitude in compactified supergravity, it
can be argued that it appears in the M theory effective action at 
one loop~\cite{StringR4}, and with the above set of derivatives at 
two loops~\cite{GreenTwoLoop}.  

The result~\eref{gravtwolooppolesB} shows that even $D=11$ supergravity
diverges.  The manifest $D$-independence of the cutting algebra allowed 
us to extend the calculation to $D=11$, even though there is no 
corresponding $D=11$ super-Yang-Mills theory.  Although we have
established a divergence, we have not given a full description of the
multiplet of counterterms, in particular how it depends on the three-form
potential $A_{\mu\nu\rho}$.  Further work on the structure of the $D=11$ 
counterterm has been carried out in ref.~\cite{DeserSeminara}.

%%%%%%%%%%%%%%%%%%%%%%%%%%%%
\subsection{Beyond two loops (a conjecture)}

Although we have not performed a full calculation beyond two loops in
either $N=4$ super-Yang-Mills theory or $N=8$ supergravity, the structure
of the entirely two-particle constructible terms leads to a natural
conjecture for where divergences appear at $L$ loops.  In the Yang-Mills
case, for each additional loop the maximum number of powers of loop
momentum in the numerator increases by two, corresponding to the insertion
of $(\ell_i+\ell_j)^2$ mentioned in section~\ref{IterationSection}.  Thus,
for $L>1$ loops we expect that the most divergent integrals have $2L - 4$
powers of loop momenta in the numerator.  There are also three additional
scalar propagators per loop.  The integrals scale as
\begin{equation}
\int (d^D p)^L  {(p^2)^{(L-2)}  \over (p^2)^{3L +1}}  \,.
\label{YMsimpInt}
\end{equation}
(The $L=1$ case is special and must be treated separately.)  
These integrals are finite for
\begin{equation}
D < {6\over L} + 4 \,,  \hskip 2 cm (L > 1) \,.
\label{N4Finiteness}
\end{equation}

The result~\eref{N4Finiteness} differs from expectations based
on $N=2$ superspace power-counting~\cite{HoweStelle}.  Specifically, for
dimensions $D=5,6$ and $7$ the amplitudes first diverge at $L = 6, 3$ and
2 loops.  The corresponding superspace arguments indicate that the first
divergence may occur at $L=4,3$ and $2$, respectively.  Although
\eref{N4Finiteness} is still only a conjecture, it is consistent with
three-loop contributions that are not detectable by the two-particle cuts,
and for which we have evaluated the three-particle cuts.  More generally,
it is also consistent with a subset of the $n$-particle cuts where
so-called `maximally helicity-violating' tree amplitudes appear on each
side of the cut~\cite{BDDPR}.  We suspect that the difference with
ref.~\cite{HoweStelle} is due to a lack of full manifest supersymmetry in
the power-counting argument.  

For $N=8$ supergravity, from insertions of $[(\ell_i+\ell_j)^2]^2$ factors
in the two-particle cuts, we expect the scaling of the most divergent
integrals at $L$ loops to be
\begin{equation}
\int (d^D p)^L {(p^2)^{2(L-2)} \over (p^2)^{3L +1}} \,. 
\label{GravsimpInt}
\end{equation}
This leads to a conjectured finiteness condition,
\begin{equation}
D < {10 \over L} + 2 \,,  \hskip 2 cm (L > 1) \,.
\label{N8Finiteness}
\end{equation}
In particular, we expect no three-loop
divergence to appear in $D=4$ --- contrary to expectations from an
$N=4$ superspace analysis~\cite{HoweStelleTownsend,HoweStelle}.
We expect the first $R^4$-type counterterm to occur at five loops instead.  
The divergence should have the same kinematical structure as the $D=7$
divergence in~\eref{gravtwolooppolesA}, but with a different
non-vanishing numerical coefficient.  

Ref.~\cite{GreenTwoLoop} argues that the coefficients of operators with
exactly four derivatives acting on $t_8t_8R^4$ in the M theory effective
action are completely accounted for by the two-loop amplitudes.  However,
the scaling behavior in~\eref{GravsimpInt}, inferred from the two-particle
cuts, suggests nonvanishing contributions of this type from all $L>2$.
A full three-loop supergravity calculation would be very welcome,
to address both the expected finiteness in $D=4$, and whether there are 
$\partial^4t_8t_8R^4$ contributions.

%%%%%%%%%%%%%%%%%%%%%%%%%%%%%%%%%%%%%%%%%%%%%%%%%%%%%%
\section{Conclusions}

The heuristic relation~\eref{GravityYMRelation} between gravity and gauge
theory is a very useful way to think about gravity, at least
perturbatively.  Unitarity allows one to bootstrap the tree-level (KLT)
versions of~\eref{GravityYMRelation} up to the multi-loop level.  Thus
gravity calculations can be performed by recycling the simpler Yang-Mills
calculations.  Although we have discussed primarily the maximally
supersymmetric case, where explicit calculations are the simplest, there
should be applications to more general cases as well.  Indeed, special
one-loop helicity amplitudes with arbitrary numbers of external gravitons
have been computed in this way, in both $N=8$ supergravity and pure
gravity~\cite{SelfDualGrav}.

We have learned that supergravity amplitudes are less divergent than
previously expected.  In $D=4$, it is still true that no supergravity
divergence has yet been firmly established.  Indeed, our expectations
in the maximally supersymmetric case, $N=8$, are for the first divergence
to occur only at five loops, not three loops.  Further work in this 
direction, using the techniques described here, could help to remedy this 
situation.

%%%%%%%%%%%%%%%%%%%%%%%%%%%%%%%%%%%

\end{document}